\documentclass[prd,aps,eqsecnum,showpacs]{revtex4}

\usepackage{amssymb}
\usepackage{bm}

\begin{document}

\title{Nonlinear curvature perturbations
in an exactly soluble model 
of multi-component slow-roll inflation}

\author{Misao {\scshape SASAKI}
%\footnote{E-mail:  @ yukawa.kyoto-u.ac.jp}
}

\affiliation{%         %Affiliation, neglected when [addenda] or [errata]
Yukawa Institute for Theoretical Physics, Kyoto University, 
Kyoto 606-8502, 
Japan
}

%\date{\today}

%%%%%%%%%%%%%%%%%%%%%
\begin{abstract}
Using the nonlinear $\delta N$ formalism,
we consider a simple exactly soluble model of multi-component
slow-roll inflation in which the nonlinear curvature perturbation
can be evaluated analytically.
\end{abstract}
%%%%%%%%%%%%%%%%%%%%%

%----------------------------
\hfill YITP-07-05
%----------------------------

\maketitle

%%%%%%%%%%%%%%%%%%%%%%%%%%%
%%%  Sec. I 
%%%%%%%%%%%%%%%%%%%%%%%%%%%
\section{Introduction}
\label{sec:intro}
Much attention has been paid to the possible
non-Gaussian signatures in the curvature perturbations
from inflation in recent years.
It is now understood that
the non-Gaussianity is generically too small to be detected
in single inflaton models with canonical kinetic term,
 as long as the standard slow-roll condition is
satisfied~\cite{Maldacena:2002vr,Seery:2005wm}.
Accordingly, various types of multi-component models have
been and are studied~\cite{NGmodels1,NGmodels2}.

In this short note, assuming that all the components
of the inflaton field are slow-rolling, we present
a simple exactly soluble model and derive the 
nonlinear curvature perturbation analytically by using
the $\delta N$ formalism~\cite{Sasaki:1998ug,Wands:2000dp,Lyth:2004gb}.
Although neither the model is compelling nor
the result is particularly new, the analysis
below is presented in the hope that it
may shed some light on general properties
of the non-Gaussianity from inflation.

\section{Model}
\label{sec:model}

We consider a multi-component inflaton field
with the potential
\begin{eqnarray}
V(\bm{\phi})=
V_0\exp\left[\frac{1}{2}\sum_{A=1}^{M}m_A^2\phi_A^2\right]\,,
\label{potential}
\end{eqnarray}
where $M$ is the number of components, and we use
the Planck units $\hbar=8\pi G=1$.
We mention that single-field models with a similar but more
general class of
potentials was analyzed extensively in \cite{Parsons:1995ew}.
Using $dN=Hdt$ as the time variable, 
where $H$ is the Hubble parameter, the slow-roll
equations of motion are
\begin{eqnarray}
\frac{\partial\phi_A}{\partial N}
=-\frac{\partial_AV}{3H^2}=-\frac{\partial_AV}{V}
=-m_A^2\phi_A\,,
\label{eom}
\end{eqnarray}
where the second equality follows from the 
potential-dominated Friedmann equation, $3H^2=V$.
This can be immediately solved to give
\begin{eqnarray}
\phi_A(N,\bm{\lambda})
=C_A(\bm{\lambda})\exp\left[-m_A^2N\right]\,,
\label{fsol}
\end{eqnarray}
where $\bm{\lambda}=(\lambda_1,\lambda_2,\cdots,\lambda_{M-1})$
is a set of parameters that specifies each trajectory
in the field space.

If the potential (\ref{potential}) were valid everywhere in
the field space, inflation would never end. We therefore
assume that there is a 'waterfall' in the field space near
the origin as in the hybrid inflation scenario~\cite{Linde:1993cn}.
For simplicity, we assume that inflation ends
when a trajectory reaches at an $(M-1)$-dimensional
deformed sphere $\Sigma$ given by
\begin{eqnarray}
\sum_{A=1}^Mm_A^{2q}\phi_A^2=D^2\,,
\label{waterfall}
\end{eqnarray}
where $q$ is a constant.

Now for the later convenience when we apply the $\delta N$ formalism,
we invert the direction of $N$ and set $N=0$ when inflation
ends, that is, when a trajectory
crosses the surface given by (\ref{waterfall}),
\begin{eqnarray}
N\to -N\,; \quad
\phi_A(0,\bm{\lambda})\in \Sigma\,.
\end{eqnarray}
Thus solution (\ref{fsol}) is re-expressed as
\begin{eqnarray}
\phi_A(N,\bm{\lambda})
=C_A(\bm{\lambda})\exp\left[m_A^2N\right]\,.
\label{solution}
\end{eqnarray}
Given the above solution, we may regard
 $(N,\lambda_1,\lambda_2,\cdots,\lambda_{M-1})$ as another
set of coordinates in the field space~\cite{Sasaki:1998ug}.
Then the coordinate $N$ is determined 
by the relation,
\begin{eqnarray}
D^2=\sum_{A=1}^M m_A^{2q}\phi_A^2\exp\left[-2m_A^2N\right]\,.
\label{Nrelation}
\end{eqnarray}
This implicitly gives $N$ as a function of the original
coordinates $\bm\phi$,
\begin{eqnarray}
N=N(\phi_1,\phi_2,\cdots,\phi_M)\,.
\end{eqnarray}
As an example, let us consider the case of
a two-component field. Then the solution is given by
\begin{eqnarray}
\phi_1=\frac{D}{m_1^q}\cos\lambda\exp[m_1^2N]\,,
\quad
\phi_2=\frac{D}{m_2^q}\sin\lambda\exp[m_2^2N]\,,
\label{2compsol}
\end{eqnarray}
where $\lambda$ parametrizes different
inflationary trajectories. 
We mention that a parametric expression for two
slow-roll fields in terms of $N$ and an angle between
the two fields similar to $\lambda$
was derived in~\cite{Polarski:1992dq}.
Eliminating $\lambda$ from the
above, we obtain $N$ as a function of $\phi_1$ and $\phi_2$.

\section{Curvature perturbation}
\label{sec:curvaturepert}

In the $\delta N$-formalism, the final amplitude
of the curvature perturbation on comoving slices
${\cal R}_c$ (or equivalently on uniform total density slices $\zeta$)
is given by $\delta N$, where $\delta N$ is the perturbation of the
e-folding number between the initial flat time-slice
at $t=t_*$
when the scale of interest came out of the horizon
during inflation and a comoving slice at $t=t_{\rm fin}$
during the final radiation-dominated stage
by which all the inflationary trajectories have
converged to a unique one~\cite{Sasaki:1998ug,Lyth:2004gb}.
 (We assume isocurvature perturbations are negligible.)

Then the (linear) curvature perturbation spectrum is
given by~\cite{Sasaki:1995aw}
\begin{eqnarray}
\frac{4\pi k^3}{(2\pi)^3}P_S(k)
\equiv \left\langle\zeta^2(t_{\rm fin})\right\rangle_k
=\left\langle{\delta N^2}\right\rangle_k
=\left(\frac{H}{2\pi}\right)_{t_{\rm k}}^2
\sum_A\left(\frac{\partial N}{\partial\phi_A}\right)_{t_{\rm k}}^2\,,
\end{eqnarray}
where $t_k$ is the horizon-crossing time of the
comoving wavenumber $k$. Here we have assumed that
there is no additional contribution to $\delta N$
after inflation. If there is such a contribution
as in the curvaton scenario~\cite{NGmodels1},
it should be added to the above. It is also noted that the tensor
perturbation has the amplitude $8H^2/(2\pi)^2$
so that the tensor-to-scalar ratio is given by~\cite{Sasaki:1995aw}
\begin{eqnarray}
\frac{P_T}{8P_S}=\frac{1}
{\displaystyle
\sum_A\left(\frac{\partial N}{\partial\phi_A}\right)_{t_{\rm k}}^2}\,.
\end{eqnarray}
Thus, the observational constraint that the left-hand side is 
more or less smaller than unity~\cite{WMAP3y} implies
\begin{eqnarray}
\sum_A\left(\frac{\partial N}{\partial\phi_A}\right)_{t_{\rm k}}^2
\gtrsim1\,.
\end{eqnarray}

The partial derivatives
$\partial N/\partial\phi_A$ can be evaluated 
from equation~(\ref{Nrelation}) as
\begin{eqnarray}
\frac{\partial N}{\partial\phi_A}=
\frac{m_A^{2q}\phi_A\exp[-2m_A^{2q}N({\bm\phi})]}
{\sum_Bm_B^{2q+2}\phi_B^2\exp[-2m_B^2N({\bm\phi})]}\,,
\label{partialN}
\end{eqnarray}
where $N$ in the right-hand side of the equation
is to be regarded as a function of $\bm{\phi}$.
For the purpose of evaluating the spectrum, it is 
more convenient to express the above in terms of
$N$ and $\bm\lambda$. Noting equations~(\ref{solution}) and
(\ref{Nrelation}), we find
\begin{eqnarray}
\frac{\partial N}{\partial\phi_A}
=\frac{n_A(\bm{\lambda})m_A^{q}\exp[-m_A^2N]}
{D\sum_B n_B^2(\bm{\lambda})m_B^2}\,;
\quad
\sum_An_A^2=1\,,
\label{multidN}
\end{eqnarray}
where $n_A(\bm{\lambda})$ is a unit vector which determines
the trajectory.
For the two-component case, using the
solution given by equation~(\ref{2compsol}), the above reduces to
\begin{eqnarray}
\frac{\partial N}{\partial\phi_1}
=\frac{m_1^{q}\cos\lambda
\exp[-m_1^2N]}{D\left(m_1^2\cos^2\lambda+m_2^2\sin^2\lambda\right)}\,,
\quad
\frac{\partial N}{\partial\phi_2}
=\frac{m_2^{q}\sin\lambda
\exp[-m_2^2N]}{D\left(m_1^2\cos^2\lambda+m_2^2\sin^2\lambda\right)}\,.
\label{2compdN}
\end{eqnarray}

The nonlinear $\delta N$ is simply given by
\begin{eqnarray}
\delta N=N(\bm{\phi}+\delta\bm{\phi})-N(\bm{\phi})\,,
\end{eqnarray}
where $\delta\bm{\phi}$ is the field perturbation
on the flat slice at the horizon-crossing time $t=t_*$.
Taking partial derivatives of equation~(\ref{partialN})
with respect to $\phi_A$ repeatedly, 
the nonlinear $\delta N$ can be easily evaluated to
an arbitrary order,
\begin{eqnarray}
\delta N=\sum_{n\geq1}\frac{1}{n!}
\frac{\partial^n N}
{\partial\phi_{A_1}\partial\phi_{A_2}\cdots\partial\phi_{A_n}}
\delta\phi_{A_1}\delta\phi_{A_2}\cdots\delta\phi_{A_n}\,.
\label{nldeltaN}
\end{eqnarray}

Since it is not particularly illuminating to spell out
the explicit formulae for higher partial derivatives of $N$,
we do not give them here. Instead let us make a rough
order of magnitude evaluation of them. 
The CMB observations~\cite{COBEDMR,WMAP3y} tell us that
\begin{eqnarray}
\left(\frac{H}{2\pi}\right)^2
\sum_A\left(\frac{\partial N}{\partial\phi_A}\right)^2\sim10^{-10}\,.
\end{eqnarray}
Thus typically we have
\begin{eqnarray}
\frac{\partial N}{\partial\phi_A}
=O\left(\frac{e^{-m^2N}}{m^2\phi_f}\right)
\sim 10^{-5}H^{-1}\,,
\label{magdN}
\end{eqnarray}
where $m$ and $\phi_f$ are
the typical magnitudes of 
$m_A$ and $\phi_A$ ($A=1,2,\cdots,M$), respectively,
such that $m^q\phi_f=D$.
It should be noted that $m^2$ must be smaller than unity
for the slow-roll condition to hold (see equation~(\ref{eom})).

For the evaluation of the order of magnitude of
higher derivatives, we may take the derivatives
of equation~(\ref{multidN}) by ignoring the dependence of
$\bm\lambda$ on $\bm\phi$. Then we obtain
\begin{eqnarray}
\frac{\partial^n N}{\partial\phi^n}
\sim m^{2n-2}
\left(\frac{\partial N}{\partial\phi}\right)^n\,.
\end{eqnarray}
Thus if the magnitude of $m^2(\partial N/\partial\phi)$ is greater
than unity, the higher derivatives become
larger than the first derivative $\partial N/\partial\phi$.
Nevertheless, if we quantify the non-Gaussianity by the 
magnitude of the $n$th order term relative to
the $n$th power of the linear term, 
such as the $f_{\rm NL}$ parameter
for the bi-spectrum~\cite{Komatsu:2001rj}, we find
\begin{eqnarray}
\zeta_{(n)}\sim
H^n\frac{\partial^n N}{\partial\phi^n}
\sim m^{2n-2}
H^n\left(\frac{\partial N}{\partial\phi}\right)^n
\sim m^{2n-2}(\zeta_{\rm (1)})^n\,,
\end{eqnarray}
where $\zeta_{\rm (1)}$ stands for the
linear part of the curvature perturbation
and $\zeta_{\rm (n)}$ for the $n$th 
order part. Thus, we may conclude that the non-Gaussianity
cannot be large unless either
there exists a field component that violates 
the slow-roll condition
or one invokes a mechanism that works
after inflation.

Finally, it may be worthwhile to point out the fact
that while $\delta N$ is formulated in Fourier space in the
linear case, it is formulated in real space in the nonlinear case~\cite{Lyth:2004gb}.
Thus $\delta\bm{\phi}$ in equation~(\ref{nldeltaN}) is given in real
space, $\delta\bm{\phi}=\delta\bm{\phi}(t,\bm{x})$,
and $t_*$ should be taken to be an epoch at which
all the relevant scales are outside the Hubble horizon.
This means $\delta\bm{\phi}$ may be affected by a small but non-trivial
effect of evolution on superhorizon scales.
Assuming the linear evolution equations are still valid for
 $\delta\bm{\phi}$, we have
\begin{eqnarray}
\delta\phi_A(t_*)=
\int\frac{d^3k}{(2\pi)^{3/2}}
\left(\hat a_{\bm k}\varphi^A_k(t_*)e^{i{\bm k}\dot{\bm x}}+h.c.\right)\,,
\end{eqnarray}
where $\hat a_{\bm k}$ is the annihilation operator for
the Bunch-Davis vacuum and
$\varphi^A_k(t)$ is the positive frequency function
as usual, and
\begin{eqnarray}
\varphi^A_k(t_*)=\frac{(\partial\phi_A/\partial N)(t_*)}
{(\partial\phi_A/\partial N)(t_k)}\varphi^A_k(t_k)\,;
\quad
\frac{4\pi k^3}{(2\pi)^3}|\varphi^A_k(t_k)|^2=
\left(\frac{H}{2\pi}\right)_{t_k}^2\,.
\end{eqnarray}
For cosmological scales of interest which span only a few
e-foldings, however, this effect may be negligible
in most cases.

\acknowledgements
This work was supported by JSPS Grant-in-Aid for
Scientific Research (S) No.~14102004, (B) No.~17340075,
 and (A) No.~18204024.

%%%%%%%%%%%%%%%%%%%%%%%%%%%%%%%%%
%% thebibliography environment 
%%%%%%%%%%%%%%%%%%%%%%%%%%%%%%%%%


\begin{thebibliography}{18}

%\cite{Maldacena:2002vr}
\bibitem{Maldacena:2002vr}
  J.~M.~Maldacena,
  % ``Non-Gaussian features of primordial fluctuations in single field
  %inflationary models,''
  JHEP {\bf 0305}, 013 (2003)
  [arXiv:astro-ph/0210603].
  %%CITATION = ASTRO-PH 0210603;%%

%\cite{Seery:2005wm}
\bibitem{Seery:2005wm}
  D.~Seery and J.~E.~Lidsey,
  %``Primordial non-gaussianities in single field inflation,''
  JCAP {\bf 0506}, 003 (2005)
  [arXiv:astro-ph/0503692].
  %%CITATION = ASTRO-PH 0503692;%%

\bibitem{NGmodels1}
%non-gaussian models
% Curvaton model
%\cite{Bartolo:2003jx}
%\bibitem{Bartolo:2003jx}
  N.~Bartolo, S.~Matarrese and A.~Riotto,
  %``On non-Gaussianity in the curvaton scenario,''
  Phys.\ Rev.\ D {\bf 69}, 043503 (2004)
  [arXiv:hep-ph/0309033].
  %%CITATION = HEP-PH 0309033;%%
\\
%\cite{Gordon:2003hw}
%\bibitem{Gordon:2003hw}
  C.~Gordon and K.~A.~Malik,
  %``WMAP, neutrino degeneracy and non-Gaussianity constraints on  isocurvature
  %perturbations in the curvaton model of inflation,''
  Phys.\ Rev.\ D {\bf 69}, 063508 (2004)
  [arXiv:astro-ph/0311102].
  %%CITATION = ASTRO-PH 0311102;%%
\\
%\cite{Enqvist:2005pg}
%\bibitem{Enqvist:2005pg}
  K.~Enqvist and S.~Nurmi,
  %``Non-gaussianity in curvaton models with nearly quadratic potential,''
  JCAP {\bf 0510}, 013 (2005)
  [arXiv:astro-ph/0508573].
  %%CITATION = ASTRO-PH 0508573;%%
\\
%\cite{Lyth:2005ab}
%\bibitem{Lyth:2005ab}
  D.~H.~Lyth,
  %``Some recent work on the curvaton paradigm,''
  Nucl.\ Phys.\ Proc.\ Suppl.\  {\bf 148}, 25 (2005).
  %%CITATION = NUPHZ,148,25;%%
\\
%\cite{Malik:2006pm}
%\bibitem{Malik:2006pm}
  K.~A.~Malik and D.~H.~Lyth,
  %``A numerical study of non-gaussianity in the curvaton scenario,''
  JCAP {\bf 0609}, 008 (2006)
  [arXiv:astro-ph/0604387].
  %%CITATION = ASTRO-PH 0604387;%%
\\
%\cite{Sasaki:2006kq}
%\bibitem{Sasaki:2006kq}
  M.~Sasaki, J.~Valiviita and D.~Wands,
  %``Non-gaussianity of the primordial perturbation in the curvaton model,''
  Phys.\ Rev.\ D {\bf 74}, 103003 (2006)
  [arXiv:astro-ph/0607627].
  %%CITATION = ASTRO-PH 0607627;%%
\\
%\cite{Valiviita:2006mz}
%\bibitem{Valiviita:2006mz}
  J.~Valiviita, M.~Sasaki and D.~Wands,
  %``Non-Gaussianity and constraints for the variance of perturbations in the
  %curvaton model,''
  arXiv:astro-ph/0610001.
  %%CITATION = ASTRO-PH 0610001;%%

\bibitem{NGmodels2}
% Inhomogeneous reheating models
%\cite{Zaldarriaga:2003my}
%\bibitem{Zaldarriaga:2003my}
  M.~Zaldarriaga,
  %``Non-Gaussianities in models with a varying inflaton decay rate,''
  Phys.\ Rev.\ D {\bf 69}, 043508 (2004)
  [arXiv:astro-ph/0306006].
  %%CITATION = ASTRO-PH 0306006;%%
\\
%\cite{Felder:2006cc}
%\bibitem{Felder:2006cc}
  G.~N.~Felder and L.~Kofman,
  %``Nonlinear inflaton fragmentation after preheating,''
  arXiv:hep-ph/0606256.
  %%CITATION = HEP-PH 0606256;%%
\\
%\cite{Alabidi:2006wa}
%\bibitem{Alabidi:2006wa}
  L.~Alabidi and D.~Lyth,
  %``Curvature perturbation from symmetry breaking the end of inflation,''
  JCAP {\bf 0608}, 006 (2006)
  [arXiv:astro-ph/0604569].
  %%CITATION = ASTRO-PH 0604569;%%

%\cite{Lyth:2004gb}
\bibitem{Lyth:2004gb}
  D.~H.~Lyth, K.~A.~Malik and M.~Sasaki,
  %``A general proof of the conservation of the curvature perturbation,''
  JCAP {\bf 0505}, 004 (2005)
  [arXiv:astro-ph/0411220].

%\cite{Sasaki:1998ug}
\bibitem{Sasaki:1998ug}
  M.~Sasaki and T.~Tanaka,
  %``Super-horizon scale dynamics of multi-scalar inflation,''
  Prog.\ Theor.\ Phys.\  {\bf 99}, 763 (1998)
  [arXiv:gr-qc/9801017].
  %%CITATION = GR-QC 9801017;%%

%\cite{Wands:2000dp}
\bibitem{Wands:2000dp}
  D.~Wands, K.~A.~Malik, D.~H.~Lyth and A.~R.~Liddle,
  % ``A new approach to the evolution of cosmological perturbations on large
  %scales,''
  Phys.\ Rev.\ D {\bf 62}, 043527 (2000)
  [arXiv:astro-ph/0003278].
  %%CITATION = ASTRO-PH 0003278;%%

%\cite{Parsons:1995ew}
\bibitem{Parsons:1995ew}
  P.~Parsons and J.~D.~Barrow,
  %``Generalized Scalar Field Potentials And Inflation,''
  Phys.\ Rev.\  D {\bf 51}, 6757 (1995)
  [arXiv:astro-ph/9501086].
  %%CITATION = PHRVA,D51,6757;%%

%\cite{Linde:1993cn}
\bibitem{Linde:1993cn}
  A.~D.~Linde,
  %``Hybrid inflation,''
  Phys.\ Rev.\ D {\bf 49}, 748 (1994)
  [arXiv:astro-ph/9307002].
  %%CITATION = ASTRO-PH 9307002;%%

%\cite{Polarski:1992dq}
\bibitem{Polarski:1992dq}
  D.~Polarski and A.~A.~Starobinsky,
  %``Spectra of perturbations produced by double inflation with an intermediate
  %matter dominated stage,''
  Nucl.\ Phys.\  B {\bf 385}, 623 (1992).
  %%CITATION = NUPHA,B385,623;%%

%\cite{Sasaki:1995aw}
\bibitem{Sasaki:1995aw}
  M.~Sasaki and E.~D.~Stewart,
  %``A General Analytic Formula For The Spectral Index Of The Density
  %Perturbations Produced During Inflation,''
  Prog.\ Theor.\ Phys.\  {\bf 95}, 71 (1996)
  [arXiv:astro-ph/9507001].
  %%CITATION = ASTRO-PH 9507001;%%

\bibitem{WMAP3y}
%\bibitem{Spergel:2006hy}
  D.~N.~Spergel {\it et al.},
 %``Wilkinson Microwave Anisotropy Probe (WMAP) three year results:
  %Implications for cosmology,''
  arXiv:astro-ph/0603449.
  %%CITATION = ASTRO-PH 0603449;%%

\bibitem{COBEDMR}
%\bibitem{Smoot:1992td}
  G.~F.~Smoot {\it et al.},
  %``Structure in the COBE differential microwave radiometer first year maps,''
  Astrophys.\ J.\  {\bf 396}, L1 (1992).
  %%CITATION = ASJOA,396,L1;%%
\\
%\bibitem{Bennett:1996ce}
  C.~L.~Bennett {\it et al.},
  %``4-Year COBE DMR Cosmic Microwave Background Observations: Maps and Basic
  %Results,''
  Astrophys.\ J.\  {\bf 464}, L1 (1996)
  [arXiv:astro-ph/9601067].
  %%CITATION = ASTRO-PH 9601067;%%

%\cite{Komatsu:2001rj}
\bibitem{Komatsu:2001rj}
  E.~Komatsu and D.~N.~Spergel,
  %``Acoustic signatures in the primary microwave background bispectrum,''
  Phys.\ Rev.\ D {\bf 63}, 063002 (2001)
  [arXiv:astro-ph/0005036].
  %%CITATION = ASTRO-PH 0005036;%%
\end{thebibliography}
\end{document}